# The effect of Joule heating on the terahertz radiation in the superconductor media

Ladan Sheikhhosseinpour ,[1] Mehdi Hosseini[1a] and Ali Moftakharzadeh [2]

[1]*Department of physics, Shiraz University of Technology, Shiraz, 313-71555, Iran*

[2]*Department of electrical engineering, Yazd University, Yazd, 89195-741, Iran*

Moving vortex lattice in type-II superconductors can result in the radiation of electromagnetic waves in the range of terahertz frequency. The vortex flow dynamics in superconductors follow the London equation. In this paper, the effects of the superconductor coherence length and the London penetration depth on the radiated power is investigated by solving the London equation in the presence of vortices. The results show that by decreasing the coherence length, the radiation power will significantly increase. Also, it is possible to obtain more radiated power with a sharp peak by increasing the penetration depth. Further investigation of the effect of Joule heating reveals that radiation power is almost independent of temperature when we are far from critical temperature. But on the other hand, near the critical temperature, the radiation power is strongly dependent on the temperature variation. Also, by increasing the bias current, the radiation power will augment, and the frequency of radiated peaks will change accordingly.

---

[a] hosseini@sutech.ac.ir

# I. INTRODUCTION

Magnetic flux penetration in a type II superconductor will result in the formation of Abrikosov vortices. The vortices manage to arrange in a triangular flux-line lattice, which is very stable. Each vortex will carry one quantum of magnetic flux $w_0 = h/2e$ [1]. Coherent movement of these fluxons inside the superconductor can radiate terahertz waveform, which is first reported by Fiory [2].

Bulaevskii and Chudnovsky fully formulated the theoretical model of electromagnetic radiation from vortex flow [3]. They showed that the superconductor radiate terahertz waves at specific washboard frequencies. The radiated power is proportional to the square of the vortex lattice velocity and the square of the magnetic field.

Because of many applications of terahertz waves in various areas like medical imaging [4–6], security systems [7,8] and communication [9,10], terahertz sources are very valuable. Semiconductor devices like resonant tunneling diode [11–14] and quantum cascade lasers [15–18] are sources of terahertz radiation. Stacked intrinsic Josephson junctions of layers of high-$T_c$ BSCCO superconductor is reported to be another source of terahertz radiation [19–24]. The total integrated radiation power of one layer of BSCCO is estimated to be 5 microwatts, and the radiated peak frequency is at 0.63 THz [24].

By exciting the special cavity mode of a stack of 580 junctions and dimensions of 300×180× 0.9 μm, the generation of 40 microwatts of coherent emission power at 0.45 THz at 80 Kelvin is reported [20]. We can manipulate the radiation peaks by adding an ac component to a dc biased superconductor. For small values of AC current, a THz modulator can be achieved. On the other hand, the addition of large values of AC component will result in a tunable frequency by adjusting the amplitude and frequency of the imposed current [25,26]. Also, the peak of radiation power will enhance by increasing the amplitude of the ac component.

Under high bias current condition or high external magnetic field, the Joule heating effect can change the superconductivity parameters, which will result in a severe variation of radiation power [27–30]. In this paper, the variation of radiated power under the influence of Joule heating in a layer of Bi-2212 is investigated.

## II. THEORETICAL MODEL

The penetrated magnetic field into a type II superconductor, forming vortices, follows the equation (1).

$$\nabla \times \nabla \times B + \frac{1}{\lambda^2} B = \frac{\Phi_0 \mathbf{e}_z}{\lambda^2} \sum_n \int dz F(r_\perp(t)) u(z) \Theta[-x_n(t)] \Theta[L_x + x_n(t)] \quad (1)$$

where $F$ is the distribution function of the magnetic field vortices, $\mathbf{e}_z$ is the unit vector along the z-axis, $\Theta(x)$ is a unit step function, $\lambda$ is the London penetration depth, $\Phi_0$ is a quantum of magnetic flux and $a = \sqrt{\Phi_0/B} \ll \lambda$ is the space between flux-lines [3].

The joule heating will increase the device temperature which will result in an increment of coherence length ($\xi$). For investigation of the widening effect of the coherence length, the vortices distribution function is assumed to be a Gaussian function as follows:

$$F(x,y) = \frac{1}{\left(\xi\sqrt{\pi}\right)^2} e^{-\frac{\vec{R}^2}{\xi^2}}, \quad \vec{R} = (x,y) - (x_n, y_p) \quad (2)$$

where $\xi$ is extremely localized for very low temperatures and can be approximated by a delta function. The equation of moving vortices in a superconductor is presented in (3):

$$x_n = an + vt, \quad y_p(z) = ap + \delta_p(z) \quad (3)$$

Here $n$, $p$ are integers and $\delta_p(z)$ accounts for disorder in the position of vortices along the y-axis.

Replacing equation (3) in equation (1) and taking the Fourier transform, we can find the resulting magnetic field from the presence of vortices in superconductors as:

$$B_{\mathcal{E}}(Š,K) = \left(K^2 + \frac{1}{\}^2}\right)^{-1} \frac{W_0}{\}^2} \sum \int dz \int e^{-iŠt} dt \int e^{-ik \cdot r} e^{-\frac{R^2}{\cdot 2}} d^3r \; \Theta\left(-\frac{an}{v} - t\right) \quad (4)$$

Taking the inverse Fourier transform of equation (4) in $k_x$ space, we will find the magnetic field at boundaries of the superconductor (x=0) [25–31]:

$$B_{\mathcal{E}}(Š, x=0, k_{y,z}) = \frac{W_0 v}{b\}} \sum_{n=1}^{N} e^{-y_p^2/\cdot^2 (2y_p/\cdot^2 - ik_y)^2 \cdot 4/4} e^{b/\}(an + 4/\cdot^2)} \int dz e^{ik_z z} \frac{e^{-i(Š - bv/\})(an/v)}}{-iŠ + ibv/\} - v}, \quad (5)$$

where $b = \left[1 + (k_y y)^2 + (k_z z)^2\right]^{0.5}$.

The total magnetic field is equal to:

$$B_z(t,r) = B_{zv}(t,r) + B_{z0}(t,r) \quad (6)$$

where $B_{z0}(t,r)$ is the solution of equation (1) with zero right hand side.

Using the superconductor radiation power equation at the frequency of [3] and replacing it in equation (5), the resulting radiation power is:

$$P_{rad}(Š) = L_y L_z \frac{k_Š W_0^2}{32 f c} \int \frac{dk_\perp}{f^2} \frac{1}{\sqrt{k_Š^2 - k_\perp^2}} S(k_\perp) \sum_{n=1}^{N} \frac{v}{b\}} e^{-<^2 b/\}(an + \cdot^2 b/(4\}))} \times \frac{e^{-iŠan/v} e^{-iban/\}}}{-iŠ + ibv/\} - v} \quad (7)$$

In which $S(k_\perp)$ is a function of vortices lattice structure and is equal to:

$$S(k_\perp) = \sum_p \int \frac{dz}{a} e^{ik_z z + y_p^2/\cdot^2 (ik_y y_p) - y_p^4/\cdot^4 + k_y^2 y_p^2/4} \quad (8)$$

here the radiation power is calculated considering strong disorder regime ($k\, al_{y,z} \ll k\, L_{y,z} \ll 1$). Using this approximation one can show that $\int dk/(2f)\, S(k) = 1/a$ and so for power we have:

$$P_{rad}(Š) = \frac{L_y L_z l_y l_z}{32 f c} \left(\frac{Ba^2 Š^2}{c}\right)^2 |\Xi(Š)|^2$$

$$\Xi(Š) = \sum_{n=1}^{\infty} \frac{v}{b\}} e^{-b/\}(an + \cdot^2 b/(4\}))} \frac{e^{-iŠan/v} e^{-iban/\}}}{-iŠ + ibv/\} - v} \quad (9)$$

because $k_\perp\} \ll 1$, therefore $b$ 1.

## III. RESULTS AND DISCUSSION

In this work, we consider Bi-2212 superconductor with $\lambda_0$=200 nm, $\xi_0$=10 nm and $T_c$=96 K. Also, in the calculations, we assume that the number of layers is 100 (N=100) and B=1 T, so a=45.5 nm [27].

The normalized radiation power versus frequency for normalized temperatures (with respect to the critical temperature) of $t$=0.98 and $t$=0.2 are depicted in Fig. 1. At frequencies near or above the superconductor energy gap, the radiation will be disrupted or disappeared entirely. Therefore, only the first ten radiation peaks have been shown in this figure. The radiation will occur at specific washboard frequencies, so the radiation power has a discrete spectrum in which its intensity will enhance by increasing frequency. This figure also indicates that the radiation peaks at a higher temperature have narrower bandwidth.

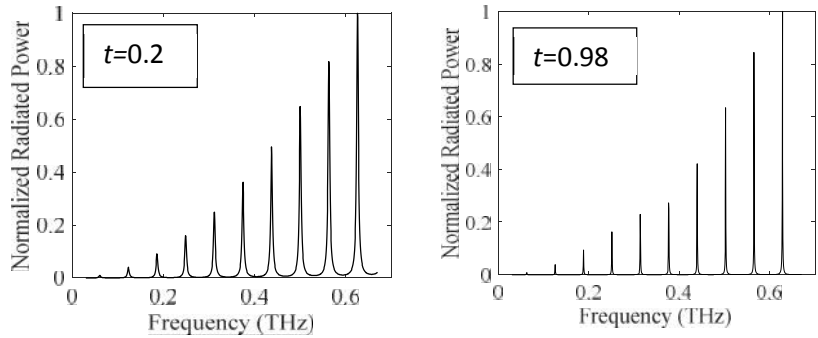

Fig1. Normalized radiated power versus frequency for normalized temperatures $t$=0.2 and $t$=0.98.

In Fig. 2, the effects of penetration depth and coherence length on the radiation power have been shown. In Fig. 2.a, by increasing the superconductor coherence length, the height of the radiation peak will reduce, but its width does not change. Because by increasing the coherence length, the superconductivity will weaken and so the radiation power will reduce.

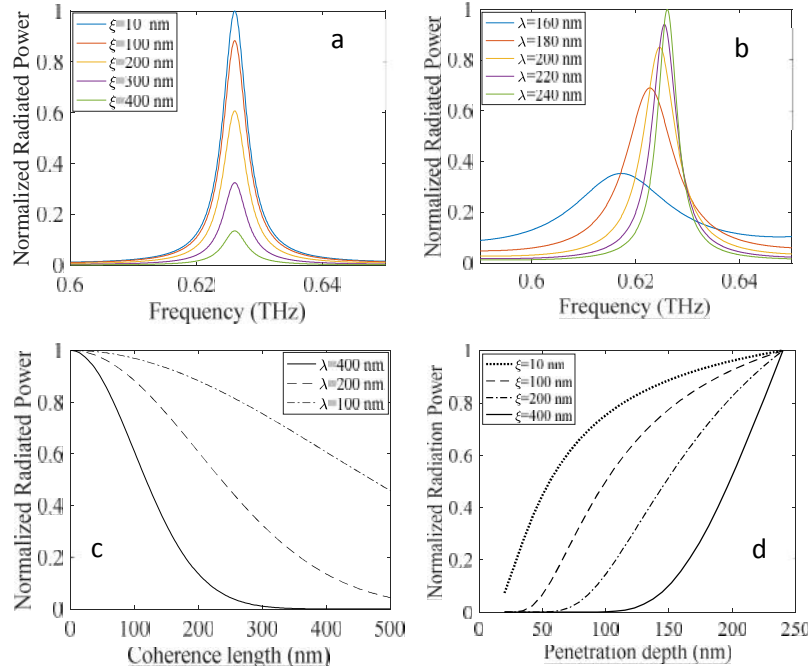

Fig. 2. Effect of   and   on normalized radiated power.

Fig. 2.b shows that we will have more radiation power for higher penetration depth. Furthermore, it is clear that by increasing the penetration depth, the width of radiation power will sharply decrease and the radiation peak will shift toward higher frequencies, this shift is negligible when we increase the coherence length. This is due to the fact that the number of vortices which can penetrate inside the superconductor without destroying its superconductivity will enhance by increasing the penetration depth. In other words, the superconductor can withstand higher magnetic fields.

Also, we can have radiated terahertz waves of higher frequency by increasing the penetration depth. Therefore, sharp and intense radiation peaks can be achieved by reducing the coherence length and enhancing the penetration depth.

Fig. 2.c shows the peak of radiation power versus the coherence length for different penetration depths. As it is obvious, the peak of radiation power will be severely reduced by increasing the coherence length, and this reduction of the peak will be augmented by reducing the penetration

depth.

The normalized radiation powers versus the penetration depth for various coherence lengths have been depicted in Fig. 2.d.

The superconductivity is powerful when the coherence length is short, so as it is shown, the radiation power will reach its maximum value at small penetration depths. On the other hand, by increasing the coherence length and so weakening the superconductivity, the rate of power enhancement is slower and begins at higher penetration depths.

Fig. 3 shows the normalized radiation power versus frequency for normalized temperatures of 0.2, 0.8, 0.98, and 0.996. As it is shown in this figure, the radiated power is broader in lower temperatures, and its peak is higher. But by approaching the critical temperature, the peak location will move to the higher frequencies. In the vicinity of the critical temperature, the power peak will decline, and also there will be some fluctuations in the radiated power.

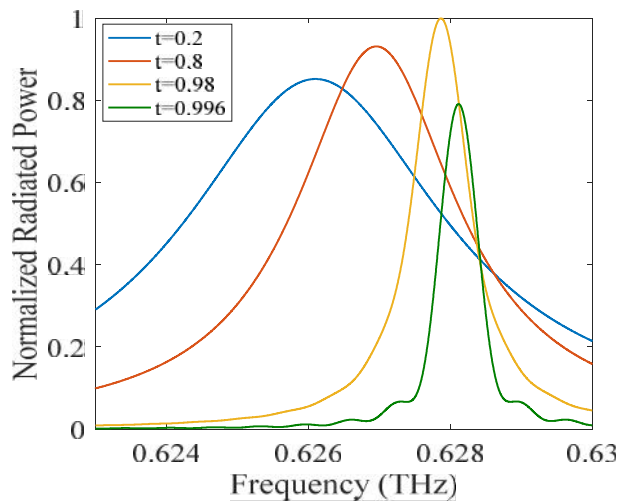

Fig. 3. Normalized radiated power versus frequency for various normalized temperature.

Fig. 4 presents the normalized radiation power versus the normalized temperature. This figure shows that the rate of change in the power intensity negligible for temperatures far below the critical temperature. Until near the critical temperature, the rate of change will enhance and, the radiated power will reach its maximum at the vicinity of the critical temperature. Just before $T_c$, the radiation power will decline rapidly and will be zero at the critical temperature.

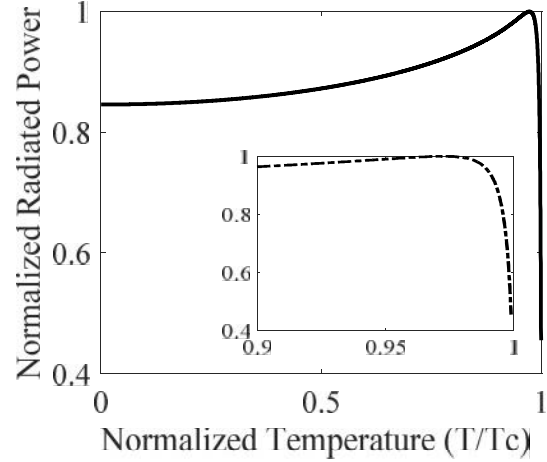

Fig. 4. Normalized radiated power versus normalized temperature.

Imposing the external current in the presence of magnetic field will generate heat in the superconductor due to the Joule heating effect [27-30]. This generated heat will enhance device resistance and will increase the temperature furthermore.

The relation between the bias current and the temperature of the superconductor device is calculated to consider the joule heating effect (Eq. 10). The physical parameters are considered as Ref. [28].

$$T = T_0 + \frac{R}{Ah}I^2 \qquad (10)$$

The critical temperature of the investigated superconductor in this work is 96 K, and the value of Ah is 2000 K/mW, in which A is the cross-section of the input flux and h is the heat transfer coefficient. As can be seen in the inset of the Fig. 5.b, increasing the bias current will result in enhancement of the temperature inside the superconductor. At the critical temperature of 96 K, the value of current is 3.1 A, and that is the critical current, beyond which the superconductivity will be destroyed, and the magnetic flux will completely penetrate in the device.

Fig. 5.a shows the normalized radiation power versus the frequency for three different currents

below the critical current. The radiation will begin by imposing the current, and it will be enhanced by increasing the current. Also, the distance between the radiated harmonics will increase. Approaching the critical current, the radiation power will reduce, and it will reach zero value at the critical current (Fig. 5.b).

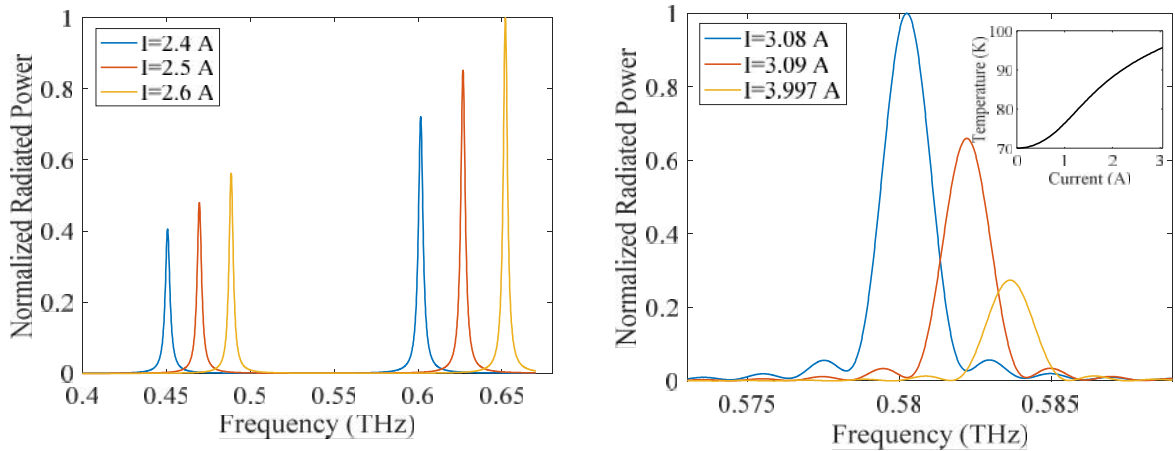

Fig.5. Normalized radiated power versus frequency for a) currents of 2.4, 2.5 and 2.6 A and b) near to critical current 3.08, 309 and 3.097A.

The behavior of normalized radiation power versus current is presented in Fig. 6. The radiated power will increase by enhancing the current, and it will reach its maximum value just before the critical current. As it is shown in Fig. 5, simultaneously with increasing the current and radiation power, the distance between the harmonics will increase, and so the peaks of radiated power will move outside the superconductor gap region frequency, therefore in Fig. 6, the radiated power will not increase monolithically and has step-like behavior.

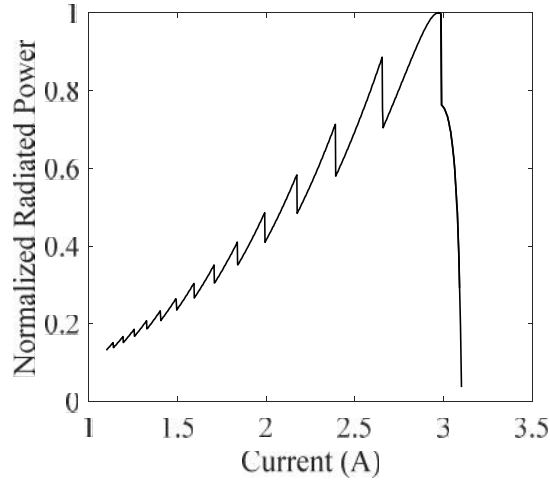

Fig. 6. Normalized radiated power versus current.

## IV. CONCLUSION

By considering the joule heating effect, the terahertz radiation from a superconductor mesa structure is investigated. The temperature and the bias current of the device are related together because of the Joule heating effect. Therefore, the effect of the device temperature on the superconductor terahertz radiation is also investigated. The Physical parameters of a superconductor such as the penetration depth and the coherence length are dependent on the device temperature. Therefore, changing the device temperature will affect the radiation power. The results show that by increasing the temperature, the radiation will increase monolithically to its maximum value near the superconductor critical temperature, and then quickly decreases to zero at the critical temperature.

The device current will affect the terahertz radiation directly and also by changing the device temperature due to the Joule heating effect. The effect of the device current on the radiated power is calculated by considering these two mechanisms. The results show that by increasing the current, the radiation power will generally increase and will reach its maximum near the critical current and then will decrease quickly.